\documentclass[preprint,aps,showpacs,nofootinbib]{revtex4}

\usepackage{epsfig,amssymb,amsmath}

\newcommand{\beq}{\begin{equation}}
\newcommand{\eeq}{\end{equation}}
\newcommand{\bea}{\begin{eqnarray}}
\newcommand{\eea}{\end{eqnarray}}
\newcommand{\bear}{\begin{array}}
\newcommand {\eear}{\end{array}}
\newcommand{\bef}{\begin{figure}}
\newcommand {\eef}{\end{figure}}
\newcommand{\bec}{\begin{center}}
\newcommand {\eec}{\end{center}}

\def\EQ#1{Eq.~(\ref{#1})}

\newcommand{\sla}[1]{\not  #1}
\begin{document}
\draft
\tighten
\preprint{DESY 13-235, TU-951, IPMU13-0226}
\title{\large \bf
Moduli-induced Baryogenesis
}
\author{
    Koji Ishiwata\,$^{a}$\footnote{email: koji.ishiwata@desy.de},
    Kwang Sik Jeong\,$^{a}$\footnote{email: kwangsik.jeong@desy.de},
    Fuminobu Takahashi\,$^{b,c}$\footnote{email: fumi@tuhep.phys.tohoku.ac.jp}
    }
\affiliation{
    $^a$ Deutsches Elektronen Synchrotron DESY, Notkestrasse 85,
         22607 Hamburg, Germany\\
    $^b$ Department of Physics, Tohoku University, Sendai 980-8578, Japan\\
    $^c$ Kavli IPMU, TODIAS, University of Tokyo, Kashiwa 277-8583, Japan
    }
%\date{\today}

\vspace{2cm}

\begin{abstract}
  We study a scenario for baryogenesis in modular cosmology and
  discuss its implications for the moduli stabilization mechanism and
  the supersymmetry (SUSY) breaking scale.  If moduli fields dominate
  the Universe and decay into the standard model particles through
  diatonic couplings, the right amount of baryon asymmetry can be
  generated through CP violating decay of gluino into quark and squark
  followed by baryon-number violating squark decay. We find that, in
  the KKLT-type moduli stabilization, at least two non-perturbative
  terms are required to obtain a sizable CP phase, and that the
  successful baryogenesis is possible for the soft SUSY breaking mass
  heavier than ${\cal O}(1)$\,TeV. A part of the parameter space for
  successful baryogenesis can be probed at the collider experiments,
  dinucleon decay search experiment, and the measurements of electric
  dipole moments of neutron and electron. It is also shown that
  similar baryogenesis works in the case of the gravitino- or the
  saxion-dominated Universe.
\end{abstract}
\pacs{}
\maketitle

%\tableofcontents
%\newpage

\section{Introduction}

The origin of baryon asymmetry is one of the most profound puzzles in
modern cosmology and particle physics. It is well-known that baryon
accounts for about 5\% of the total energy density of the
Universe~\cite{Ade:2013zuv}:
\begin{eqnarray}
n_B/s = (8.6 \pm 0.1) \times 10^{-11}~~ (68\%~{\rm C.L.}),
\label{eq:YBobs}
\end{eqnarray}
where $n_B$ is the number density of baryon, and $s$ is the entropy
density.  The standard model (SM) of particle physics and standard
cosmology, however, do not provide any viable mechanism to generate
the observed baryon number. To be specific, three conditions known as
the Sakharov conditions~\cite{Sakharov:1967dj} have to be satisfied
for successful baryogenesis: (i) baryon number violation, (ii) C and
CP violation, and (iii) departure from thermal equilibrium.  These
conditions require physics beyond the SM and set tight constraints on
possible scenarios for baryogenesis.

The mechanism for baryogenesis is closely related to the thermal
history of the Universe.  The recent observations of cosmic microwave
background as well as large scale structure of the Universe firmly
support the inflationary Universe in which our Universe experienced
an accelerating cosmic expansion at an early stage of the
evolution~\cite{Guth:1980zm,Linde:1981mu}.  Thus baryon asymmetry must
be generated after inflation since otherwise the exponential expansion
during inflation would completely dilute any pre-existing baryon
asymmetry.

However, the history of the Universe after inflation is poorly
known. The inflaton may directly decay into the SM particles to reheat
the Universe. The right amount of baryon asymmetry can be generated
through thermal leptogenesis if the reheating temperature is
sufficiently high~\cite{Fukugita:1986hr}. On the other hand, the
evolution of the Universe could be more involved. Indeed, there are
many moduli fields in supergravity and superstring theories, which may
have a significant effect on the thermal history. They are known to
easily dominate the energy density of the Universe because the moduli
fields are copiously produced by the coherent oscillation and that
their interactions are extremely weak, suppressed by the Planck scale
or Grand Unified Theory (GUT) scale.  Then a huge amount of entropy is
released by the moduli decay, diluting pre-existing baryon asymmetry.
As a consequence we have two possibilities for baryogenesis. One is to
create a sufficiently large amount of baryon asymmetry before the
moduli decay by, {\it e.g.}, the Affleck-Dine
mechanism~\cite{Affleck:1984fy,Dine:1995kz}, which has been
extensively studied in a context of modular
cosmology~\cite{Stewart:1996ai,Asaka:1999xd,Kasuya:2001tp,Jeong:2004hy,Kawasaki:2007yy,Higaki:2012ba}.
The other is to generate baryon asymmetry after the moduli decay. This
is the main focus of the present paper.

We assume that moduli fields dominate the energy density of the
Universe in the framework of supergravity. After dominating the
Universe, moduli decay to lighter particles in the minimal extension
of supersymmetric standard model (MSSM).  The moduli decay temperature
can be higher than ${\cal O}(1)$~MeV for the moduli mass heavier than
about $10$~TeV if their interactions are suppressed by the Planck
mass, thus avoiding bounds from the big bang nucleosynthesis
(BBN)~\cite{LowTRH}.  In this paper we study a baryogenesis scenario
in which the baryon asymmetry is generated through CP violating decay
of gluino into quark and squark, followed by baryon-number violating
squark decay.  We will introduce renormalizable R-parity violating
interactions to get the required baryon number violation. Late-time
moduli decay can easily realize the out-of-equilibrium decays of
gluino and squark. On the other hand, the size of the CP phase depends
on how supersymmetry (SUSY) breaking is transmitted to the visible
sector.  As we will see shortly, the required CP violation for the
baryogenesis has important implications for moduli stabilization.
Successful baryogenesis is possible in moduli stabilization such that
moduli-mediated SUSY breaking is sizable, and the axionic shift
symmetry associated with the modulus is broken at least by two terms
in the modulus superpotential.  We will focus on mixed modulus-anomaly
mediation in the Kachru-Kallosh-Linde-Trivedi (KKLT)-type moduli
stabilization~\cite{Kachru:2003aw}, as it provides a natural framework
for the moduli-induced baryogenesis.  We will also show that similar
baryogenesis works in other cases where the gravitino, the saxion, or
the Polonyi field dominates the Universe, and decays mainly into the
SM gauge sector.

Let us mention differences of the present paper from the works in the
past.  Baryogenesis induced by a modulus-like particle has been
studied since long time ago.  The gravitino-induced baryogenesis was
proposed by Cline and Raby~\cite{Cline:1990bw}, followed by Mollerach
and Roulet who studied a similar scenario where the saxion plays the
role of the gravitino~\cite{Mollerach:1991mu}.  Recently baryogenesis
via such late-decaying particles was studied in a generic way by with
higher dimension operators~\cite{Cheung:2013hza}.  In those studies
the origin of the CP phase was simply assumed to be independent of the
decaying particle.  In the present paper we study a baryogenesis
within a concrete moduli stabilization mechanism, where the modulus
field mediates SUSY breaking with a non-vanishing CP phase, and its
decay provides out-of-equilibrium decays of gluino and squark.
Therefore the moduli-induced baryogenesis studied in this paper offers
a realistic and self-contained scenario.  Later in this paper we will
revisit the gravitino-induced baryogenesis.  Here we will provide
analytic formulae which can be applied to a more generic mass spectrum
of superparticles.  Note that it was shown that the
inflaton~\cite{Kawasaki:2006gs,Asaka:2006bv,Endo:2006qk,Endo:2007ih,Endo:2007sz}
as well as moduli~\cite{Endo:2006zj} generically decays into a pair of
gravitinos. Therefore the gravitino-dominated Universe can be realized
in a broader scenario than originally assumed. Its cosmological and
phenomenological implications have been studied recently in
Ref.~\cite{Jeong:2012en}.  We will also revisit the saxion-induced
baryogenesis. In Ref.~\cite{Mollerach:1991mu}, it was indicated that
the saxion decay to gauginos is helicity-suppressed. As clearly shown
in Ref.~\cite{Endo:2006ix}, however, it is not helicity-suppressed and
its rate is generically comparable to that into gauge bosons. Taking
into account this fact, we will show that the saxion-induced
baryogenesis is more efficient than originally considered before.

The rest of this paper is organized as follows. In Sec.~\ref{sec:2} we
study mixed modulus-anomaly mediation in the KKLT-type moduli
stabilization, focusing on how to obtain the CP violation required for
successful baryogenesis.  We will then examine the baryogenesis by
modulus decay in Sec.~\ref{sec:3}, and study the implications for the
SUSY breaking scale and discuss its experimental consequences.  In
Sec.~\ref{sec:4} we discuss similar baryogenesis by the decay of the
gravitino and the saxion. Sec.~\ref{sec:5} is devoted to conclusions
and discussion.  In this paper we take the reduced Planck mass
$M_P\simeq 2.4 \times 10^{18}~{\rm GeV}$ to be unity unless otherwise
noted.

%%%%%%%%%%%%%%%%%%%%%%%%%%%%%%%%%
\section{Moduli stabilization}
\label{sec:2}
%%%%%%%%%%%%%%%%%%%%%%%%%%%%%%%%%

We consider the possibility that late-decaying moduli account for the
observed baryon asymmetry of the Universe.  To produce baryon
asymmetry, we need couplings that violate both baryon number and CP.
Baryon number violation arises from R-parity violating interactions
while satisfying various experimental constraints, such as
neutron-antineutron oscillation, dinucleon decay, and proton decay.
The required CP violation, on the other hand, can be induced after
SUSY breaking from the R-parity violating interactions.  However,
generating a nonzero CP phase is nontrivial.  It depends on the SUSY
breaking mechanism and the moduli stabilization.

As is well-known, anomaly mediation~\cite{anomaly.mediation} (and also
gauge mediation) generates flavor and CP conserving soft SUSY breaking
terms because it takes place mainly through the SM gauge interactions.
The soft terms induced by moduli $F$-terms are also considered to
preserve both flavor and CP.  Here the flavor conservation is a result
of flavor-universal rational numbers called the modular weight, which
determines the coupling between moduli and the matter fields in
visible sector~\cite{dilaton.mediation}, meanwhile CP conserving soft
terms are the consequence of axionic shift symmetries associated with
the moduli \cite{Choi:1993yd}.  However, this is true only for a
simple KKLT-type moduli stabilization.  It turns out that a sizable CP
violation can be obtained from the moduli sector if
\begin{itemize}
\item Soft SUSY breaking terms receive sizable contributions from the
  moduli $F$-terms, and
\item Non-perturbative corrections, which generate a modulus potential
  and break the axionic shift symmetry, involve at least two terms.
\end{itemize}
We will see that the KKLT scenario is a natural framework for the
moduli-induced baryogenesis that works at low temperatures.  It will
be also shown in the next section that such a sizable CP violation for
baryogenesis could have interesting experimental consequences.

\subsection{KKLT mechanism}

In the KKLT compactification \cite{Kachru:2003aw}, all the moduli are
stabilized at a phenomenologically viable vacuum.  There are three
types of moduli; dilaton, complex structure moduli, and K\"ahler
moduli. The dilation and complex structure moduli are fixed by fluxes
while obtaining masses hierarchically larger than the gravitino mass.
Incorporating non-perturbative corrections to the superpotential, one
can also stabilize K\"ahler moduli at a supersymmetric anti-de Sitter
(AdS) minimum.  Since our vacuum is a SUSY breaking de Sitter with a
tiny cosmological constant, we introduce an anti-brane stabilized at
the tip of a highly warped throat. Then its tension provides a small
positive energy to cancel the negative cosmological constant.  At the
same time it explicitly breaks the $N=1$ SUSY preserved by the
background geometry and fluxes. To be precise, a small vacuum shift is
induced and makes the moduli develop nonzero $F$-terms.  Through
Planck-suppressed interactions with the moduli, sparticles then obtain
the soft SUSY breaking masses as
\begin{eqnarray}
\label{moduli-F}
m_{\rm soft} \sim \frac{m^2_{3/2}}{(\mbox{modulus mass})},
\end{eqnarray}
where $m_{3/2}$ denotes the gravitino mass.  Hence, if stabilized with
mass much larger than $4\pi^2 m_{3/2}$, the modulus does not play an
important role in generating the soft SUSY breaking terms because
they always receive anomaly-mediated SUSY breaking contribution,
$m_{\rm soft}\sim m_{3/2}/4\pi^2$, in supergravity \cite{anomaly.mediation}.

The stabilization of K\"ahler moduli can be examined within the
effective theory after integrating out the dilation and complex
structure moduli as they are much heavier than the gravitino and
K\"ahler moduli~\cite{Choi:2004sx}. Let us consider a simple case with
a single K\"ahler modulus $X$.\footnote{ With an abuse of notation, we
  shall use the same symbol to denote both a chiral superfield and its
  scalar component, unless noted otherwise.  } It is straightforward
to generalize to a case with multi K\"ahler moduli.  The K\"ahler
potential takes the no-scale form at the leading order in the
$\alpha^\prime$ (the string tension) and the string loop expansions,
\begin{eqnarray}
K_0 &=& -3\ln(X+X^*).
\label{eq:K0}
\end{eqnarray}
It should be noticed that the theory possesses the axionic shift
symmetry,
\begin{eqnarray}
{\rm U(1)}_X: \quad {\rm Im}(X)\to {\rm Im}(X) + {\rm constant}.
\label{eq:pert_shift_sym}
\end{eqnarray}
This shift symmetry can be explicitly broken at non-perturbative
level. For example, hidden gaugino condensation or stringy instanton
effects induce non-perturbative terms in superpotential. Including the
non-perturbative corrections, the modulus superpotential is given by
\begin{eqnarray}
W_0 &=& \omega_0 + W_{\rm np}(X),
\label{eq:W0}
\end{eqnarray}
where the constant term $\omega_0$ is originated from background
fluxes, which is assumed to be much smaller than order unity in the
Planck unit so as to get the soft SUSY breaking scale much lower than
the Planck scale.  We also note that the constant term explicitly
breaks U$(1)_R$.  The above superpotential stabilizes the modulus at a
SUSY AdS minimum satisfying
\begin{eqnarray}
\label{susy-condition}
\partial_X W_0 +(\partial_X K_0)W_0 =0.
\end{eqnarray}
Here $\partial_X$ means a partial derivative with respect to the
scalar field $X$, {\it i.e.}, $\partial_X \equiv
\frac{\partial}{\partial X}$.  Then the modulus mass is determined by
the curvature around the minimum as
\begin{eqnarray}
m_X \simeq \left| \left\langle
\frac{\partial_X K_0}{\partial^2_X K_0}
\frac{\partial^2_X W_{\rm np}}{\partial_X W_{\rm np}} \right\rangle
 m_{3/2}
\right|,
\label{eq:mX_0}
\end{eqnarray}
where the gravitino mass is given by $m_{3/2}=\langle e^{K/2}W \rangle
\sim \omega_0$, and we have used $\partial_X K_0=\partial_{X^*} K_0$
and the SUSY condition (\ref{susy-condition}). After adding a
sequestered uplifting potential $V_{\rm lift}\propto e^{2K_0/3}$ to
cancel the negative cosmological constant, a small vacuum shift is
induced. This vacuum shift results in a nonzero modulus $F$-term,
\begin{eqnarray}
\label{F-term}
\langle F^X \rangle \simeq
-\frac{2m^*_{3/2}}{\langle \partial^2_X W_{\rm np}/\partial_X W_{\rm np} \rangle},
\end{eqnarray}
which is order of $|m_{3/2}|^2/m_X$.

Let us continue to examine how the SUSY breaking is mediated to the
visible sector.  The soft SUSY breaking terms receive contributions
from both modulus mediation and anomaly mediation
\cite{Choi:2004sx,Endo:2005uy,Choi:2005uz},
\begin{eqnarray}
-{\cal L}_{\rm soft} =
m^2_i |\phi_i|^2
+ \left(\frac{1}{2}M_a \lambda_a\lambda_a
+ A_{ijk}y_{ijk}\phi_i\phi_j\phi_k
 + {\rm h.c.} \right),
\label{eq:Lsoft}
\end{eqnarray}
where $\phi_i$ denotes the scalar component of the visible sector
chiral superfield $\Phi_i$, $\lambda_a$ is the gaugino, and $y_{ijk}$
is the Yukawa coupling for the superpotential term
$\Phi_i\Phi_j\Phi_k$.  Here $i$, $j$, $k$ are the indices of matter
fields, and $a=1,2,3$ corresponds to the U(1)$_Y$, SU(2)$_L$, SU(3)$_C$
gauge group, respectively.  The mass of gaugino is obtained as
\begin{eqnarray}
\label{gaugino-mass}
M_a(\Lambda) &=&
-\langle F^X\partial_X \ln g^2_a(\Lambda) \rangle
+ \frac{b_ag^2_a(\Lambda)}{16\pi^2}\,m^*_{3/2}
\nonumber \\
&=&
\left\langle \frac{F^X}{X+X^*} \right\rangle
+ \frac{b_ag^2_a(\Lambda)}{16\pi^2}\,m^*_{3/2},
\end{eqnarray}
 when the gauge kinetic function is 
\begin{eqnarray}
f_a(\Lambda) = k_a X,
\label{eq:fa}
\end{eqnarray}
at the cut-off scale $\Lambda$, which we assume to be the GUT scale.
$k_a$ is a positive constant.  Here we note that the gauge kinetic
function is allowed to have only a term which is linear to $X$ (except
for constant term) due to the axionic shift symmetry, {\it i.e.},
Eq.~(\ref{eq:pert_shift_sym}).  The gauge coupling reads
$g^2_a=1/\langle {\rm Re}(f_a) \rangle$, and the beta-function
coefficients are given by $b_a=(33/5,1,-3)$ in the MSSM.  On the other
hand, $m_i$ and $A_{ijk}$ depend on the anomalous dimension $\gamma_i$
and the modular weight $n_i$ of $\phi_i$~\cite{Kaplunovsky:1993rd}
that determines the modulus dependence of the matter wave function
$Z_i$.\footnote{ The expression of the scalar soft SUSY breaking terms
  can be found, for instance, in Ref.~\cite{Choi:2005uz}.  }  For
instance, the $A$-parameter reads
\begin{eqnarray}
\label{A-parameter}
A_{ijk}(\Lambda) &=&
\langle F^X\partial_X \ln(Z_iZ_jZ_k) \rangle
- \frac{\gamma_i(\Lambda)+\gamma_j(\Lambda)+\gamma_k(\Lambda)}{16\pi^2}
\,m^*_{3/2}
\nonumber \\
&=&
(n_i+n_j+n_k) \left\langle \frac{F^X}{X+X^*} \right\rangle
- \frac{\gamma_i(\Lambda)+\gamma_j(\Lambda)+\gamma_k(\Lambda)}{16\pi^2}
\,m^*_{3/2},
\end{eqnarray}
where $n_i$ is a rational number of order unity, and it can have
various values in the KKLT-type moduli stabilization with anomalous
U(1) gauge symmetry \cite{Choi:2006bh}.  Low energy values of the soft
SUSY breaking parameters are determined by the renormalization group
running, and extra gauge-charged matters if they exist at an
intermediate scale \cite{Everett:2008qy,Choi:2009jn,Nakamura:2008ey}.

\subsection{CP violation in the moduli sector}

As will be shown in the next section, in order to generate sufficient
baryon asymmetry from modulus decays, the phase of the combination
$A_{ijk}M^*_{\tilde g}$ should be sizable, {\it i.e.},
\begin{eqnarray}
\arg(A_{ijk} M^*_{\tilde g}) \neq 0,
\end{eqnarray}
where $M_{\tilde g}$ is the gluino mass, and hereafter $A_{ijk}$
denotes the soft trilinear parameter associated with baryon-number
violating superpotential terms, which will be presented
soon.

In the KKLT scenario, moduli and anomaly mediations are comparable to
each other, and the relative phase between the modulus $F$-term and
the gravitino mass is not rotated away in the presence of two or more
non-perturbative superpotential terms.  This indeed makes it as a
natural framework to implement the baryogenesis in the
modulus-dominated Universe.  From the relations (\ref{gaugino-mass})
and (\ref{A-parameter}), the phase is naively estimated to be
\begin{eqnarray}
\label{CP-violation-for-baryogenesis}
\arg(A_{ijk} M^*_{\tilde g}) \sim
\frac{{\arg}(m_{3/2}\langle F^X \rangle )}{\alpha+\alpha^{-1}},
\end{eqnarray}
omitting an order unity coefficient which depends on the
renormalization group running and extra gauge-charged matter fields.
The parameter $\alpha$ represents the ratio between moduli and anomaly
mediations,
\begin{eqnarray}
\alpha
\equiv \left|\,
\left\langle\frac{F^X}{X+X^*}\right\rangle^{-1}
\frac{m_{3/2}}{4\pi^2}
\right|
\approx \frac{m_X}{8\pi^2 m_{3/2}},
\end{eqnarray}
where we have used that the modulus $F$-term is given by
Eq.~(\ref{F-term}).  The original KKLT scenario gives $\alpha \simeq
\ln(M_P/m_{3/2})/4\pi^2={\cal O}(1)$.  It is obvious that the phase is
suppressed if one of the mediation mechanisms dominates over the
other.  On the other hand, the approximate symmetry, U$(1)_R$ and
U$(1)_X$, indicates that $W_{\rm np}$ should include at least two
non-perturbative terms to get nonzero $\arg(m_{3/2}\langle F^X
\rangle)$.  This is understood as follows.  Suppose that there is only
a single non-perturbative term.  One can then remove the CP phases in
the superpotential by redefining $\omega_0$ and $X$.  However, it is
not generically possible to rotate away the CP phase(s) of additional
non-perturbative term(s). That is why at least two non-perturbative
terms are needed to get a CP phase.\footnote{The mirage unification
  pattern of gaugino masses~\cite{Endo:2005uy,Choi:2005uz} is violated
  in our baryogenesis scenario where the CP violation results from a
  relative phase between the modulus F-term and the gravitino mass.}

Now let us estimate the CP phase more quantitatively.  For this
purpose, we consider the superpotential,
\begin{eqnarray}
\label{W-model}
W_0 = \omega_0 - Ae^{-a X} - Be^{-b X},
\end{eqnarray}
with $a\neq b$ and $a\sim b \sim 4\pi^2$, where $A$ and $B$ are order
unity complex constants.  Here the approximate U$(1)_R$ and U$(1)_X$
(explicitly broken by the constant and non-perturbative terms) allow
us to make $\omega_0$ and $A$ be real and positive numbers without
loss of generality, which is the convention we will take hereafter.
This explains why $X$ is fixed at a CP-conserving minimum with mass
\begin{eqnarray}
m_X \simeq 2 m_{3/2} \ln(M_P/m_{3/2}),
\label{eq:mX}
\end{eqnarray}
in the absence of the last term in Eq.~(\ref{W-model}), as was
considered in the original KKLT mechanism. Here we have used
$m_{3/2}\sim \omega_0$ and $A\sim 1$.  Let us now turn on the
$Be^{-bX}$ term with ${\rm Im}(B)\neq 0$.  Then the SUSY
condition reads
\begin{eqnarray}
a Ae^{-aX} + b Be^{-bX}  \simeq
\frac{3 \omega_0}{ X+X^* },
\end{eqnarray}
neglecting small terms suppressed by $1/a(X+X^*)$ or $1/b(X+X^*)$.
Since the right-hand side is a real number, ${\rm Im}(B)\neq 0$ shifts
the minimum to a CP-violating one, {\it i.e.}, $\langle {\rm Im}(X)
\rangle\neq0$.  Using the facts that the modulus $F$-term is generated
according to the relation (\ref{F-term}) after uplifting, and that the
modulus is fixed at $a \langle {\rm Re}(X) \rangle \simeq
\ln(M_P/m_{3/2})$, we find that $\arg(m_{3/2}\langle F^X \rangle)$ is
determined by
\begin{eqnarray}
\arg(m_{3/2}\langle F^X\rangle) \simeq
-{\rm arg}\left\langle
\frac{\partial^2_X W_{\rm np}}{\partial_X W_{\rm np}}\right\rangle,
\end{eqnarray}
where the superpotential (\ref{W-model}) gives
\begin{eqnarray}
\label{Wxx}
\frac{\partial^2_X W_{\rm np}}{\partial_X W_{\rm np}}
= -a - (b-a)
\frac{b Be^{-bX}}{a Ae^{-aX} + b Be^{-bX}}.
\end{eqnarray}
Therefore, the CP phase can be sizable if $aAe^{-aX}$ and $bBe^{-bX}$
are comparable to each other in size.\footnote{ One may consider a
  racetrack-type model \cite{Krasnikov:1987jj} where the modulus is
  stabilized mainly by the competition between two non-perturbative
  terms in the superpotential while the constant piece is negligibly
  small.  However, in this case, two approximate symmetries U$(1)_R$
  and U$(1)_X$ fix the modulus near a CP preserving minimum.
  Furthermore, the modulus becomes much heavier than
  $m_{3/2}\ln(M_P/m_{3/2})$, thereby suppressing the modulus
  mediation.  See the relation (\ref{CP-violation-for-baryogenesis}).
}  Under the assumption that $|bBe^{-bX}|\lesssim |aAe^{-aX}|$, the
vacuum shift induced by ${\rm Im}(B)$ leads to
\begin{eqnarray}
\label{CP-in-KKLT}
\arg(m_{3/2}\langle F^X\rangle)
&\approx&
\frac{b(b-a)}{a^2}
\left( \frac{m_{3/2}}{{M_P}}\right)^{\frac{b-a}{a}} \frac{{\rm Im}(B)}{A},
\end{eqnarray}
for $b$ similar to but larger than $a$.  Here $a/b$ is a rational
number if the non-perturbative terms arise from hidden gaugino
condensation.  For instance, the phase is around $0.05$--$0.1$
$m_{3/2}\sim 10^4$--$10^8~{\rm GeV}$ in the model with $(b-a)/a= 1/10$
for $A\sim 0.1$ and ${\rm Im}(B)\sim 1$.

Let us close this section by briefly mentioning the SUSY breaking
scale.  We are interested in the case where the CP violation in the
moduli sector gives rise to a sizable $\arg(A_{ijk} M^*_{\tilde g})$
to implement the baryogenesis. Thus, to avoid the SUSY CP problems,
the MSSM sparticles should be heavier than 1~TeV unless there is a
cancellation among sparticle contributions in the amplitudes of
physical processes. We will quantitatively discuss the constraint on
the CP phase later. Heavy sparticles around over TeV scale may
indicate a little hierarchy problem regarding the Higgs boson
mass. However, to put it another way, such a large sparticle (mainly
stop) mass is one of possible explanations for the 126~GeV Higgs
boson discovered at the LHC~\cite{Aad:2012tfa,Chatrchyan:2012ufa}.

%%%%%%%%%%%%%%%%%%%%%%%%%%%%%%%%%
\section{Baryogenesis}
\label{sec:3}
%%%%%%%%%%%%%%%%%%%%%%%%%%%%%%%%%

In this section we will discuss the generation of baryon asymmetry by
modulus decay. During inflation the modulus is likely deviated from
the true vacuum in the low energy, due to the deformation of the
potential through gravitational interactions with the inflaton.  After
inflation ends, it starts coherent oscillation when the Hubble
parameter becomes comparable to $m_X$ with a large initial
displacement of the order of the Planck scale, and then soon dominates
the energy density of the Universe.  Eventually the modulus field
decays to lighter particles. The modulus decay releases a huge amount
of entropy, which would dilute any (harmful) relic of the early
Universe, and reheats the Universe at the temperature $T_X$ given by
\begin{eqnarray}
T_X \simeq \left(\frac{90}{\pi^2 g_*(T_X)} \right)^{1/4}
\sqrt{\Gamma_X M_{P}},
\label{eq:TR}
\end{eqnarray}
where $g_*$ is the relativistic degrees of freedom, and $\Gamma_X$ is
the total decay width of the modulus field.  Although the late modulus
decay might washout pre-existing baryon asymmetry as well, we will
show that sufficient baryon asymmetry is generated by the modulus
decay followed by the gluino decay if the R-parity is violated.  Here
we will also discuss the experimental consequences of this
baryogenesis scenario. 

\subsection{Baryon asymmetry}
\label{sec:Basym}

The modulus decays into the MSSM particles through its couplings to the
visible sector.  For the case that the gauge couplings are determined
by $\langle X \rangle$, the modulus $X$ dominantly decays into gauge
boson pairs, and gaugino pairs through the
interactions (written in two-component notation)~\cite{Endo:2006zj},
\begin{eqnarray}
{\cal L}_{X} = \frac{c^a_g}{4}\left( \delta X_r  G_{\mu\nu}^aG^{a\mu\nu}
- \delta X_i  G_{\mu\nu}^a \tilde G^{a\mu\nu} \right)
- \frac{c^a_\lambda}{4} \left(  \delta X_r \lambda_a \lambda_a -
i \delta X_i \lambda_a \lambda_a
+ {\rm h.c.} \right),
\end{eqnarray}
with $\delta X_r + i \delta X_i \equiv \langle 2 \partial^2_X K_0
\rangle^{1/2} (X - \langle X \rangle)$ being the canonically
normalized fluctuation about the vacuum.  Here $G_{\mu\nu}^a$ is the
field strength of gauge field, and the modulus couplings are given by
\begin{eqnarray}
c^a_g &=&
\sqrt{2} \,\frac{\partial_X \ln
    g^2_a}{\sqrt{\partial^2_X K_0}},
  \nonumber \\
c^a_\lambda &=& c^a_g \left(
1+ {\cal O}\Big(\frac{m_{3/2}}{m_X}\Big) \right) m_X,
\end{eqnarray}
where the modulus coupling to gauginos has been derived by using the relation
\begin{eqnarray}
\partial_X F^{X^*} = -e^{K_0/2}\frac{\partial^2_X W_0}{\partial^2_X K_0}\left(
1+ {\cal O}\Big(\frac{m_{3/2}}{m_X}\Big) \right),
\end{eqnarray}
evaluated at the vacuum. Then the decay rates to a gauge boson pair and
a gaugino pair are given by
\begin{eqnarray}
&&
\Gamma(X \rightarrow A^aA^a)= \frac{1}{96\pi} \frac{m_X^3}{M_P^2},
\\ &&
\Gamma(X \rightarrow \lambda_a\lambda_a)=
\frac{1}{96\pi}\frac{m_{X}^3}{M_P^2}
\left[1-\frac{4M_{a}^2}{m_X^2}\right]
\left[1-\frac{5M_{a}^2}{2m_X^2}\right],
\end{eqnarray}
respectively, for both the real and imaginary components of $X$.  Here
$A^a$ denotes the gauge boson, neglecting its mass, and we have used
the gauge kinetic function $f_a=k_a X$.  Under the assumption that
modulus decays into hidden sector particles are suppressed, the total
decay rate of the modulus is simply written as
\begin{eqnarray}
\Gamma_X \simeq \frac{1}{4\pi} \frac{m^3_X}{M^2_P},
\label{eq:Gamma_X}
\end{eqnarray}
neglecting the gaugino mass in the final state.
Then, from Eqs.~(\ref{eq:TR}) and (\ref{eq:Gamma_X}), the
reheating temperature is estimated as
\begin{eqnarray}
T_X \simeq  98~{\rm GeV}
\left(\frac{g_\ast}{106.75}\right)^{-1/4}
\left(\frac{m_X}{10^8~{\rm GeV}}\right)^{3/2}.
\label{eq:TRvalue}
\end{eqnarray}
Since the decay rate is Planck-suppressed, the reheating temperature
is usually lower than the typical SUSY breaking scale, which means
that the produced sparticles are out of equilibrium.  This is an
essential ingredient, {\it i.e.}, for the Sakharov condition (ii), for
the baryogenesis via subsequent gluino and squark decays. It is also
important that the branching fraction into a gluino pair is given by
\begin{eqnarray}
{\rm Br}({X\rightarrow \tilde g\tilde g})
=\frac{\Gamma({X\to \tilde g\tilde g})}{\Gamma_X}
\simeq \frac{1}{3}.
\label{eq:Brgl}
\end{eqnarray}
Thus gluinos, whose decay is the source of the baryon asymmetry, are
abundantly produced by modulus decay.

For the baryogenesis we introduce the baryon-number violating
renormalizable operators in addition to the MSSM superpotential
$W_{\rm MSSM}$;\footnote{ Instead, we can consider other R-parity
  violating operators which break lepton number. In such a case,
  nonzero lepton number would be generated in a similar manner as is
  illustrated in the following discussion.  Then the lepton number is
  converted to baryon number via the sphaleron
  process. See Ref.~\cite{Krauss:2013dia}. }
\begin{eqnarray}
W_{\rm vis}=W_{\rm MSSM} + \frac{1}{2} \lambda_{ijk} U^c_i D^c_j D^c_k,
\end{eqnarray}
where $U^c_i$ and $D^c_j$ are the SU(2)$_L$ singlet up-type and
down-type quarks, respectively, and $i,j,k$ are flavor indices while
color indices are implicit. The other renormalizable lepton number
violating operators can be forbidden by generalized lepton parities or
discrete R symmetries~\cite{Barbier:2004ez}.  From the superpotential
we obtain the following Lagrangian,
\begin{eqnarray}
{\cal L}_{\sla{R}_p}&=&
-\lambda_{ijk}
\Bigl(
\tilde{d}_k^c (\bar{u}_iP_Ld^c_j) + \tilde{u}_i^c (\bar{d}_jP_Ld^c_k)
\Bigr)
-\lambda_{ijk}^*
\Bigl(
\tilde{d}_k^{c*} (\bar{d}^c_jP_Ru_i) + \tilde{u}_i^{c*} (\bar{d}^c_kP_Rd_j)
\Bigr)
\nonumber \\ &&
+\, \frac{1}{2}(A_{ijk}\lambda_{ijk}\tilde{u}_i^c\tilde{d}_j^c\tilde{d}_k^c
+A_{ijk}^*\lambda_{ijk}\tilde{u}_i^{c*}\tilde{d}_j^{c*}\tilde{d}_k^{c*}).
\label{eq:L_RPV}
\end{eqnarray}
Here the tilde denotes the scalar superpartner, and $P_{R/L}$ are
projection operators defined as $P_{R/L}=(1\pm \gamma_5)/2$ (plus for
$P_R$ and minus for $P_L$).  The first line in the right-hand side
includes SUSY couplings, while the second line is the soft SUSY
breaking trilinear terms, which correspond to the second term in
Eq.~(\ref{eq:Lsoft}).

\begin{figure}[t]
  \begin{center}
    \includegraphics[scale=0.8]{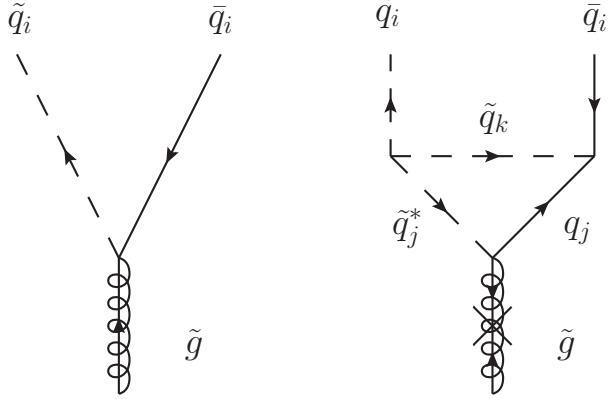}
  \end{center}
  \caption{Gluino decay diagrams to a quark and a squark.
    The cross in the gluino line represents a chirality flip.}
\label{fig:gluinodecay}
\end{figure}

The gluino mainly decays to quark and squark through the tree-level
diagram in Fig.~\ref{fig:gluinodecay}.  The decay rate for $\tilde g
\rightarrow \tilde{q}_i \bar{q}_i$ reads
\begin{eqnarray}
\Gamma(\tilde g \rightarrow \tilde{q}_i\bar{q}_i) =
\frac{\alpha_s}{8} |M_{\tilde g}|(1-r_i)^2,
\end{eqnarray}
where $r_i=m_{\tilde{q}_i}^2/|M_{\tilde g}|^2$, neglecting the quark
mass.  Here $\alpha_s=g^2_3/4\pi$ is the strong coupling constant, and
$m_{\tilde{q}_i}$ denotes the mass of squark in the final state.  The
total decay rate of the gluino is given by the sum of all possible
final sates of quark and squark,
\begin{eqnarray}
\Gamma_{\tilde g}
= \sum_i \left[\Gamma({\tilde g \rightarrow \tilde{q}_i\bar{q}_i})
+ \Gamma({\tilde g \rightarrow \tilde{q}^*_i q_i})\right].
\end{eqnarray}
On the other hand, an asymmetry between $\Gamma({\tilde g \rightarrow
  \tilde{q}_i\bar{q}_i})$ and $\Gamma({\tilde g \rightarrow
  \tilde{q}^*_i q_i})$ may be induced at the loop level. If there
exists another decay mode to $\tilde{q}_j^* q_j$, the interference
between tree and loop diagrams (Fig.~\ref{fig:gluinodecay}) generates
nonzero asymmetry via the baryon-number violating operator given in
Eq.~(\ref{eq:L_RPV});
\begin{eqnarray}
\Delta \Gamma({\tilde{g}\rightarrow \tilde{q}_{Ri}\bar{q}_{Ri}}) &\equiv&
\Gamma({\tilde g \rightarrow \tilde{q}_{Ri} \bar{q}_{Ri}}) -
\Gamma({\tilde g \rightarrow \tilde{q}^*_{Ri} q_{Ri}})
\nonumber \\
&=&
\sum_{j,k} c_{ij} \frac{\alpha_s}{32\pi}
\frac{|\lambda_{ijk}|^2{\rm Im}(A_{ijk}M^*_{\tilde g})}{|M_{\tilde g}|}
f(r_i,r_j,r_k),
\label{eq:DeltaGamma}
\end{eqnarray}
where
\begin{eqnarray}
f(r_i,r_j,r_k) = (1-r_i)(1-r_j)
 - r_k \ln \left(1+r^{-1}_k(1-r_i)(1-r_j)\right).
\label{eq:f}
\end{eqnarray}
Here the constant $c_{ij}$ has a nonzero value only for the process
kinematically allowed: $c_{ij}=2$ if the intermediate and final states
({\it i.e.}, ${\tilde q_j}^* q_j$ and ${\tilde q_i} \bar{q}_i$) are both
down-type (s)quarks, and $c_{ij}=1$ otherwise.  In the above, we have
ignored left-right mixing in the squark sector for simplicity. It is
straightforward to take into account the left-right mixing. The
function $f(r_i,r_j,r_k)$ behaves as $f(r_i,r_j,r_k)\simeq 1$ when
$r_{i,j,k}\ll 1$. On the other hand, it is possible that $r_k$ is
greater than one.  If $r_k>1$ (and $r_{i,j}\ll 1$), then it is
suppressed as $f(r_i,r_j,r_k)\simeq 1/2r_k$. This behavior is expected
because the loop diagram in Fig.~\ref{fig:gluinodecay} should vanish
in the limit $m_{\tilde{q}_k} \rightarrow \infty$.\footnote{ This
  point was not discussed in Ref.~\cite{Cline:1990bw}.}

Now we discuss net baryon number generated by the gluino decay.  First
we write the (s)quark number which is generated by the decay of a
single gluino as
\begin{eqnarray}
 \Delta n_{\tilde{q}_i} = - \Delta n_{q_i}
=
\frac{\Delta \Gamma({\tilde{g} \rightarrow \tilde{q}_{Ri} \bar{q}_{Ri}})}
{\Gamma_{\tilde g}}.
\label{eq:delanq}
\end{eqnarray}
Even though it is zero just after the gluino decay, nonzero baryon
number is generated by subsequent $\Delta B \neq 0$ squark decay
processes.  Possible decay modes of the squark are $\Delta B=1$ processes
such as $\tilde{q}_i \rightarrow \bar{q}_j \bar{q}_k$, and $\Delta
B=0$ processes such as $\tilde{q} \rightarrow
\tilde{\chi}^0q$ with $\tilde{\chi}^0$ being a neutralino.
Here the $\Delta B=1$ process also includes the case where
$\tilde{q}\rightarrow \tilde{q}'W/Z$ and subsequently $\tilde{q}'$
decays to quark pairs via the baryon-number violating operator.
Eventually the net baryon number generated by a single gluino is
obtained as
\begin{eqnarray}
\epsilon_B &\equiv& \sum_{i}
\Delta n_{\tilde{q}_i}
\left[
{\rm Br}^{\tilde{q}_i}\times\Bigl(-\frac{2}{3} \Bigr)+
(1-{\rm Br}^{\tilde{q}_i})\times \frac{1}{3}
\right]
+\sum_{i}\Delta n_{q_i} \times \frac{1}{3}
\nonumber \\
&=&-\sum_{i} \Delta n_{\tilde{q}_i} {\rm Br}^{\tilde{q}_i}.
\end{eqnarray}
Here ${\rm Br}^{\tilde{q}_i}$ is the branching fraction for the
$\Delta B=1$ decay process, which can be the dominant mode when the
$\lambda_{ijk}$ is order unity, or it is unity when ${\tilde{q}_i}$ is
the lightest sparticle. Then, the baryon number density from a gluino
decay is
\begin{eqnarray}
n_B &=& n_{\tilde g}\,\epsilon_B
\nonumber \\
&=& 2n_X {\rm Br}(X\rightarrow \tilde{g}\tilde{g}) \epsilon_B
\nonumber \\
&\simeq& \frac{2}{3} n_X \epsilon_B,
\end{eqnarray}
where $n_{\tilde{g}}$ and $n_X$ are the number density of gluino and
the modulus, respectively, and we have used Eq.~(\ref{eq:Brgl}) at the
last step.  Under the assumption that all the energy density of the
modulus turns into radiation at its decay, {\it i.e.}, $\rho_X=m_X n_X
\simeq \rho_R$ ($\rho_R$ is the energy density of radiation), the
yield of baryon asymmetry is obtained as
\begin{eqnarray}
\frac{n_B}{s}
&=& \frac{3T_X}{4m_X} 2{\rm Br}(X\rightarrow \tilde{g}\tilde{g})\epsilon_B.
\end{eqnarray}

Let us estimate the baryon asymmetry in this baryogenesis scenario.
For the estimation of the asymmetry, we assume that only
$\lambda_{332}$ is nonzero for simplicity.\footnote{Actually the case
  where only $\lambda_{332}$ is sizable is realistic in
  the phenomenological point of view. One of well-motivated scenarios is
  the minimal flavor violation~\cite{Csaki:2011ge}. }
  The total decay width of the gluino is then given by
\begin{eqnarray}
\Gamma_{\tilde g }
&\simeq& 4n_f \Gamma({\tilde g \rightarrow \tilde{q}_i \bar{q}_i})
\simeq \frac{\alpha_sn_f}{2}|M_{\tilde g}|.
\end{eqnarray}
Here a factor four counts CP conjugate, and left-handed and
right-handed fields, and $n_f$ is the number of quark flavors to which
gluino can decay.  We have taken the limit $r_i\ll 1$ for the
kinematically allowed decay for simplicity.  In the same limit and
taking $n_f=6$, then the asymmetry is given by
\begin{eqnarray}
\Delta \Gamma_{\tilde{g}} &=&
\sum_i \Delta \Gamma({\tilde{g} \rightarrow \tilde{q}_{Ri} \bar{q}_{Ri}})
\nonumber \\
 &\simeq&
 \frac{\alpha_s}{4\pi}
\frac{|\lambda_{332}|^2{\rm Im}(A_{332}M^*_{\tilde g})}{|M_{\tilde g}|},
\end{eqnarray}
which leads to\footnote{This expression has a different factor
  compared to the result given in Ref.~\cite{Cline:1990bw}. In
  Ref.~\cite{Cline:1990bw}, it is $-\frac{1}{16\pi}$ instead of
  $\frac{1}{12\pi}$. }
\begin{eqnarray}
\frac{\Delta \Gamma_{\tilde{g}}}{\Gamma_{\tilde g}}
\simeq
 \frac{1}{12\pi}
\frac{|\lambda_{332}|^2{\rm Im}(A_{332
}M^*_{\tilde g})}{|M_{\tilde g}|^2}.
\label{eq:epsilon}
\end{eqnarray}
This is the typical scale of asymmetric parameter when the gluino is
much heavier than the squarks. Furthermore
taking ${\rm Br}^{\tilde t,\, \tilde b,\, \tilde s}={\rm Br}^{\tilde{q}}$, we get
\begin{eqnarray}
\epsilon_B \simeq  1.3 \times 10^{-3}
\left(\frac{|\lambda_{332}|^2\,{\rm Im}
(A_{332} M^*_{\tilde g})/|M_{\tilde g}|^2}{-0.1}\right)
 \left(\frac{{\rm Br}^{\tilde{q}}}{0.5}\right),
\label{eq:epsilonB}
\end{eqnarray}
which results in 
\begin{eqnarray}
\frac{n_B}{s}
&\simeq&
4.9\times 10^{-10} \left(\frac{g_\ast}{106.75}\right)^{-1/4}
\left(\frac{m_X}{10^8~{\rm GeV}}\right)^{1/2}
\left(\frac{\epsilon_B}{10^{-3}}\right).
\label{eq:etaB}
\end{eqnarray}
Therefore, the observed baryon asymmetry of the Universe can be
obtained with ${\cal O}(0.1)$ CP phase and ${\cal O}(1)$ baryon-number
violating coupling if the modulus $X$ has a mass around 
$10^6$--$10^8$~GeV.  In the later numerical calculation, we
also take into account the effect of sphaleron processes.  Namely, we
replace $n_B/s$ as $(28/79)\times n_B/s$ when $T_X\gtrsim 4\pi
m_W/g_2^2$, where $m_W$ is the $W$ boson mass.

\subsection{Numerical result}

\begin{figure}[t]
  \begin{center}
    \includegraphics[scale=0.8]{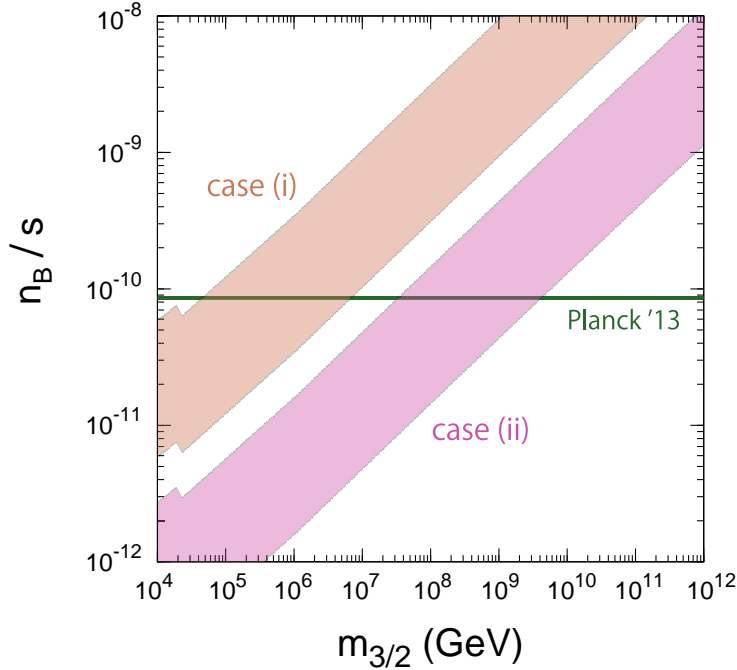}
  \end{center}
  \caption{Baryon asymmetry generated by modulus decays as a function
    of the gravitino mass for cases where $(i)$ $m_{\tilde t,\,\tilde
      b,\,\tilde s}=m_{\rm soft}$, ${\rm Br}^{\tilde t,\,\tilde
      b,\,\tilde s}=0.5$ and $(ii)$ $m_{\tilde t,\,\tilde b}=m_{\rm
      soft}$, ${\rm Br}^{\tilde t,\,\tilde b}=0.5$, where $m_{\rm
      soft}=m_{3/2}/4\pi^2$.  For both cases, the gluino mass is taken
    as $M_{\tilde{g}}=3m_{\rm soft}$, and the other squark masses are
    taken to be $m_{\tilde q}=6 m_{\rm soft}$.  For the modulus mass,
    we take $m_X=2m_{3/2}\ln(M_P/m_{3/2})$ following
    Eq.~(\ref{eq:mX}).  Upper and lower bands correspond to the baryon
    asymmetry in case $(i)$ and $(ii)$, respectively. For both cases,
    the CP phase is in the region, $0.01\le -|\lambda_{332}|^2 ({\rm
      Im}(A_{332} M^*_{\tilde g})/|M_{\tilde g}|^2) \le 0.1$.  The
    observed value given in Eq.~(\ref{eq:YBobs}) is depicted as
    ``Planck~$'$13''.}
  \label{fig:eta}
\end{figure}

In this subsection we give the numerical results.  Since we are
considering the KKLT moduli stabilization where the modulus and
anomaly mediations give comparable contributions to the soft SUSY
breaking terms, the typical mass scale for the sparticles is given by
\begin{eqnarray}
m_{\rm soft} =\frac{m_{3/2}}{4\pi^2}.
\label{eq:m_soft}
\end{eqnarray}
The precise sparticle mass spectrum is model-dependent, {\it i.e.}, it
relies on the details of moduli stabilization and moduli couplings to
the visible sector, renormalization group flow from the cut-off scale,
and presence or absence of intermediate scale gauge-charged matters.
In the following analysis we parametrize the relevant sparticle masses
by $m_{\rm soft}$, and use the relation Eq.~(\ref{eq:mX}) for the
modulus mass.  For simplicity, we take only $\lambda_{332}$ to be
nonzero.  Let us consider the following two examples:
\begin{eqnarray}
(i) &&
M_{\tilde g} = 3m_{\rm soft}, \quad m_{\tilde t,\,\tilde b,\,\tilde s}
= m_{\rm soft}, \quad
m_{\tilde q\neq \tilde t,\,\tilde b,\,\tilde s} = 6 m_{\rm soft},
\quad
{\rm Br}^{\tilde t,\,\tilde b,\,\tilde s} = 0.5,
\nonumber \\
(ii) &&
M_{\tilde g} = 3m_{\rm soft}, \quad m_{\tilde t,\,\tilde b}
= m_{\rm soft}, \quad
m_{\tilde q\neq \tilde t,\,\tilde b} = 6 m_{\rm soft}, \quad
{\rm Br}^{\tilde t,\,\tilde b} = 0.5.
\nonumber
\end{eqnarray}
For both cases we take the effective CP violation parameter to be
\begin{eqnarray}
0.01~\le~
-|\lambda_{332}|^2 \frac{{\rm Im}(A_{332} M^*_{\tilde g})}{|M_{\tilde g}|^2}
~\le~ 0.1,
\label{eq:CPphase_range}
\end{eqnarray}
where we have taken into account that the KKLT moduli stabilization
leads to CP violation according to Eq.~(\ref{CP-in-KKLT}) and assumed
$|\lambda_{332}|\sim {\cal O}(1)$.  The observed baryon asymmetry
given in Eq.~(\ref{eq:YBobs}) is drawn in the plot. The baryon
asymmetry in the case $(ii)$ is expected to be suppressed by
$M_{\tilde{g}}^2/m_{\tilde{s}}^2$ as we mentioned in the previous
subsection. We show it as an reference of one of the possible SUSY
mass spectra.

Fig.~\ref{fig:eta} shows $n_B/s$ as a function of $m_{3/2}$.  The
baryon asymmetry from the modulus decay is shown by the upper band for
the case $(i)$, and by the lower one for the case $(ii)$.  We have
checked the out-of-equilibrium condition for the gluino and the squark
is fulfilled in a wide range of the parameter space, such as $m_{3/2}
\lesssim 10^{14}~{\rm GeV}$.  One can see that, in the case $(i)$, the
correct amount of baryon asymmetry is generated when the gravitino
mass is around $10^5$--$10^7$~GeV, which corresponds to $m_{\rm
  soft}\sim 10^3$--$10^5$~GeV.  This is consistent with a rough
estimation given in Eq.~(\ref{eq:etaB}).  On the other hand, in the
case $(ii)$, the suppression by the strange squark mass makes the
baryon asymmetry smaller compared to the previous case by around an
order of magnitude.\footnote{ The suppression factor can be estimated
  as follows. Since the gluino cannot decay to strange (s)quark,
  $\Gamma_{\tilde{g}}$ becomes $(2/3)\times\Gamma_{\tilde{g}}$
  compared to the case $(i)$, and $\Delta \Gamma_{\tilde{g}}$ becomes
  $(1/4)(M_{\tilde{g}}^2/2m_{\tilde{s}}^2)\times\Delta
  \Gamma_{\tilde{g}}$. This gives a suppression factor $3/63$, which
  is consistent with the numerical result given in
  Fig.~\ref{fig:eta}.}  As a consequence, a larger gravitino mass is
required to account for the baryon asymmetry of the Universe.  The
figure shows that the gravitino mass should lie in the range about
$10^7$--$10^{10}$~GeV.

The gravitino can be produced by the modulus decay with a branching
fraction of ${\cal O}(0.01)$. The produced gravitino acts as radiation
until it decays because it is originally produced
relativistically. Thus it never dominates the Universe, and it does
not cause significant entropy production. In addition, the gravitino
decay does not destroy the successful BBN if $m_{3/2}\gtrsim 10~{\rm
  TeV}$, which is satisfied in the parameter range of our interest.
Therefore, the asymmetry generated by modulus (and subsequent
sparticle) decays is not diluted.  It is also important to note that
the moduli-induced gravitino problem \cite{Endo:2006zj} is avoided in
the presence of R-parity violating operators because the lightest
sparticle is not stable anymore.

Finally we comment on constraints from other experiments.  A large CP
phase coming from $\arg(A_{ijk} M^*_{\tilde g})$ may induce sizable
electric dipole moments (EDMs) of neutron and electron.  The current
experimental bounds are given by~\cite{Baker:2006ts,Baron:2013eja},
\begin{eqnarray}
\begin{array}{ll}
d_n \le 2.9 \times 10^{-26}~e~{\rm cm}&~~({\rm 90\%~ C.L.}),
\\ 
d_e \le 8.7 \times 10^{-29}~e~{\rm cm}&~~({\rm 90\%~ C.L.}),
\end{array}
\end{eqnarray}
for neutron and electron EDMs, respectively. We have estimated the
neutron and electron EDM based on
Refs.~\cite{Pospelov:2005pr,Hisano:2012sc,Hisano:2012cc} (see also
Refs.~\cite{Polchinski:1983zd,DeRujula:1990wy} and
Ref.~\cite{Fuyuto:2013gla}), and found that both constraints give a
similar bound as
\begin{eqnarray}
{\rm Im}(A_{ijk}M_{\tilde{g}}^*)/|M_{\tilde{g}}|^2 
\lesssim 0.1
\times \left(\frac{m_{\rm soft}}{1~{\rm TeV}}\right)^2.
\end{eqnarray}
Here we have taken the universal soft SUSY breaking mass, and
neglected the CP phase from $\mu/B\mu$ term~\cite{Choi:2005uz}.
Therefore the current bound allows ${\cal O}(0.1)$ CP phase even if
sparticles are as low as around TeV, which confirms the
parametrization in the numerical calculation, {\it
  i.e.}, Eq.~(\ref{eq:CPphase_range}).  If the CP phase from $\mu/B\mu$
term is included, the bound may become more stringent by a factor,
depending on the model. In any case, it is interesting since the
parameter space we are interested in may be probed in the ongoing or
future measurements of the EDMs.

The baryon-number violating operators possibly induce
neutron-antineutron ($n\mathchar`-\bar{n}$) oscillation or dinucleon
decay. Those experiments constrain the coupling in light flavors, such
as $ud \tilde{s}$ (in two-component notation). However, even if
$\lambda_{112}$ is negligibly small, sizable $\lambda_{332}$ induces
$ud \tilde{s}$-type coupling at one loop. For instance,
$\lambda_{332}=1$ gives a value of ${\cal O}(10^{-7})$ for the light
flavor coupling~\cite{Dimopoulos:1987rk}. Thus it may constrain our
baryogenesis scenario. Super-Kamiokande experiment gives constraint on
the time scale of $n\mathchar`-\bar{n}$ oscillation and dinucleon
decay in ${}^{16}$O~\cite{Abe:2011ky,Litos:2010zra},
\begin{eqnarray}
\begin{array}{ll}
\tau_{n\mathchar`-\bar{n}}\ge 2.4 \times 10^{8}~{\rm sec}
&~~({\rm 90\%~ C.L.}),
\\ 
\tau(pp \rightarrow K^+K^-) \ge 1.7 \times 10^{32}~{\rm yr}
&~~({\rm 90\%~ C.L.}).
\end{array}
\end{eqnarray}
We have estimated the time scale of $n\mathchar`-\bar{n}$ oscillation
and the decay rate of $pp \rightarrow K^+K^-$ based on
Refs.~\cite{Dimopoulos:1987rk,Goity:1994dq}, then we found the
experimental bounds above give rise to
\begin{eqnarray}
&&
|\lambda_{112}|\lesssim 4.4 \times 10^{-3}
\left(\frac{m_{\rm soft}}{1~{\rm TeV}}\right)^{5/6}
\left(\frac{250~{\rm MeV}}{\tilde{\Lambda}}\right),
\\ &&
|\lambda_{112}|\lesssim 3.2 \times 10^{-7}
\left(\frac{m_{\rm soft}}{1~{\rm TeV}}\right)^{5/2}
\left(\frac{250~{\rm MeV}}{\tilde{\Lambda}}\right)^{5/2},
\end{eqnarray}
respectively.  Here we have set $M_{\tilde{g}}=m_{\tilde{q}}=m_{\rm
  soft}$ for simplicity. $\tilde{\Lambda}$ is the hadronic scale used
in the evaluation of the hadronic matrix element, which has
theoretical uncertainty.  As it is seen, the dinucleon decay
experiment gives more stringent constraint than $n\mathchar`-\bar{n}$
oscillation.
%because of the flavor structure of the baryon-number violating interactions. 
Especially the bound from dinucleon decay is interesting.  Recall that
the numerical result implies that $m_{\tilde q}\sim {\cal O}(1) ~{\rm
  TeV}$ with $\lambda_{332}\sim {\cal O}(1)$ leads to the right
abundance of baryon ({\it i.e.}, case(i)).  This means that a part of
the parameter region where our baryogenesis works can be probed in the
current or future experiments. In addition, the squark with a mass of
${\cal O}(1)~{\rm TeV}$ may be discovered at the collider experiments,
such as the LHC or the ILC.

%%%%%%%%%%%%%%%%%%%%%%%%%%%%%%%%%
\section{Other scenarios}
\label{sec:4}
%%%%%%%%%%%%%%%%%%%%%%%%%%%%%%%%%

So far we have discussed baryogenesis by the modulus decay.  As it has
been shown, the KKLT scenario has important ingredients for the
baryogenesis, {\it i.e.}, suppressed interactions of the moduli with
the visible sector, a large enough CP phase and a large branching
fraction to gluinos in the modulus decay. Other than the KKLT
scenario, there are several possible scenarios which possess (some of)
the ingredients stated above.  In this section we will discuss the
gravitino- and the saxion-induced baryogenesis.

\subsection{Gravitino-induced baryogenesis}
The baryogenesis by the gravitino decay was originally discussed by
Cline and Raby in Ref.~\cite{Cline:1990bw}. In their work relatively
low SUSY breaking scale, which is around several hundred GeV, was
studied. Recently Ref.~\cite{Krauss:2013dia} pursued the
gravitino-induced leptogenesis using lepton number violating
operators. Here we revisit the original case by Cline and Raby, also
considering high scale SUSY breaking region.

The gravitino is produced effectively at a high temperature. If the
reheating temperature after inflation is high enough, the gravitino is
copiously generated, and eventually dominates the Universe.  In
addition, the gravitino can also be produced by the inflaton
decay~\cite{Kawasaki:2006gs,Asaka:2006bv,Endo:2006qk,Endo:2007ih,Endo:2007sz}.
In the typical mass spectra given in Eq.~(\ref{eq:m_soft}), the
gravitino decays to the MSSM particles.  The total decay rate is given
by
\begin{eqnarray}
\Gamma_{3/2}=
\sum_i \left[\Gamma({\psi_\mu \rightarrow \tilde{\phi}_i \phi_i})
+\Gamma({\psi_\mu \rightarrow \bar{\tilde{\phi}}_i \phi_i^*})\right]
+\sum_a \Gamma({\psi_\mu \rightarrow \lambda_a A_a}),
\end{eqnarray}
where $\psi_\mu$ is the gravitino, $\tilde{\phi}_i$ is the fermionic
superpartner of scalar $\phi_i$ and\footnote{We found a typo for
  $\Gamma({\psi_\mu \rightarrow \tilde{\phi}_i \phi_i})$ given in
  Ref.~\cite{Moroi:1995fs}.  We thank T.~Moroi for confirming this
  point. For $\Gamma({\psi_\mu \rightarrow \lambda_a A_a})$, we got
  consistent result with Ref.~\cite{Moroi:1995fs}, which is four times
  smaller than the result given in Ref.~\cite{Cline:1990bw}.}
\begin{eqnarray}
&&
\Gamma({\psi_\mu \rightarrow \tilde{\phi}_i \phi_i})
=\frac{1}{192 \pi}\frac{m^3_{3/2}}{M_P^2}
\left[1-\frac{m_{\phi}^2}{m_{3/2}^2}\right]^4,
\\ &&
\Gamma({\psi_\mu \rightarrow \lambda_a A_a})
=\frac{1}{32 \pi}\frac{m^3_{3/2}}{M_P^2}
\left[1-\frac{M_a^2}{m_{3/2}^2}\right]^3
\left[1+\frac{M_a^2}{3m_{3/2}^2}\right].
\end{eqnarray}
Here we have neglected gauge boson masses.  Since the decay rates are
Planck suppressed as in moduli decay, the reheating temperature is
typically the same as Eq.~(\ref{eq:TRvalue}). Now let us suppose that
the gravitino decays to all MSSM particles.  Then taking the large
gravitino mass limit, the total decay width is given by
\begin{eqnarray}
\Gamma_{3/2}\simeq \frac{1}{32 \pi}\left(\frac{1}{3}N+N_g\right)
\frac{m_{3/2}^3}{M_P^2}
\end{eqnarray}
where $N=49$ is the number of chiral superfields (neglecting
right-handed neutrinos) and $N_g=12$ is the number of gauginos in the
MSSM. Then the reheating temperature is estimated as
\begin{eqnarray}
T_{3/2}\simeq  1.9\times 10^2~{\rm GeV}
\left(\frac{g_\ast}{106.75}\right)^{-1/4}
\left(\frac{m_{3/2}}{10^8~{\rm GeV}}\right)^{3/2}.
\label{eq:TR_{3/2}}
\end{eqnarray}

In the gravitino decay a fair amount of gluinos is produced if it is
kinematically allowed.  The branching ratio of the decay to gluino and
gluon, under the assumption above, is then
\begin{eqnarray}
{\rm Br}(\psi_\mu \rightarrow \tilde{g}g)
=\frac{\Gamma({\psi_\mu \rightarrow \tilde{g}g})}{\Gamma_{3/2}}
\simeq \frac{8}{N/3+N_g}=\frac{24}{85}.
\label{eq:Br_psi_gluino}
\end{eqnarray}
This large branching fraction is important for the baryogenesis. Net
baryon asymmetry from the gluino decay is then expressed as
\begin{eqnarray}
\frac{n_B}{s}\Big|_{\rm gluino}  =
\frac{3T_{3/2}}{4m_{3/2}}{\rm Br}(\psi_\mu \rightarrow \tilde{g}g)\epsilon_B.
\label{eq:etaB_{3/2}}
\end{eqnarray}
Using Eqs.~(\ref{eq:TR_{3/2}}) and (\ref{eq:Br_psi_gluino}), $n_B/s$
is estimated as
\begin{eqnarray}
\frac{n_B}{s}\Big|_{\rm gluino} &\simeq&
3.9\times 10^{-10} \left(\frac{g_\ast}{106.75}\right)^{-1/4}
\left(\frac{m_{3/2}}{10^8~{\rm GeV}}\right)^{1/2}
\left(\frac{\epsilon_B}{10^{-3}}\right).
\label{eq:etaB_{3/2}_gluino}
\end{eqnarray}

On the other hand, the decay modes to quark and squark are also
important. In the mass spectrum mentioned above, the branching
fraction is
\begin{eqnarray}
{\rm Br}(\psi_\mu \rightarrow  \tilde{q}_{R}\bar{q}_{R})=
\frac{\sum_i\left[\Gamma({\psi_\mu \rightarrow \tilde{q}_{Ri}\bar{q}_{Ri}})
+\Gamma({\psi_\mu \rightarrow \tilde{q}_{Ri}^*q_{Ri}})\right]}{\Gamma_{3/2}}
\simeq \frac{3n_f}{N/3+N_g}=\frac{9n_f}{85},
\label{eq:Br_psi_squark}
\end{eqnarray}
where $n_f$ is the number of quark flavors which can be produced by
the gravitino decay.  As it is seen, the branching fraction can be
comparable to ${\rm Br}(\psi_\mu \rightarrow \tilde{g}g)$.
Note that this decay mode has the asymmetry between its CP conjugate
and itself, which is given by a similar diagram depicted in
Fig.~\ref{fig:gluinodecay} just by replacing the gluino with the
gravitino. Therefore if physical CP phase ${\rm Im}(A_{ijk}m_{3/2}^*)$
is nonzero, then baryon asymmetry is generated. As a result of
straightforward calculation, we get
\begin{eqnarray}
\Delta \Gamma({\psi_{\mu} \rightarrow \tilde{q}_{Ri} \bar{q}_{Ri}})
= \sum_{j,k}  \frac{1}{256\pi^2}
\frac{|\lambda_{ijk}|^2{\rm Im}(A_{ijk}m^*_{3/2})|m_{3/2}|}
{M_P^2}
g(r_i,r_j,r_k),
\end{eqnarray}
where
\begin{eqnarray}
g(r_i,r_j,r_k)
&=&(1-r_i)(1-r_j)\left[ (1-r_i)(1-r_j)+6r_k\right]
\nonumber \\ &&
-2r_k\left[2(1-r_i)(1-r_j)+3r_k\right]
\ln \left[1+\frac{(1-r_i)(1-r_j)}{r_k}\right].
\end{eqnarray}
Here $r_i=m_{\tilde{q}_i}^2/|m_{3/2}|^2$. Similarly to
$f(r_i,r_j,r_k)$, $g(r_i,r_j,r_k)$ approaches to unity when
$r_{i,j,k}\ll 1$, while it is suppressed as $g(r_i,r_j,r_k) \simeq
1/2r_k$ when $r_k>1$ and $r_{i,j}\ll 1$ as expected. Then net baryon
number directly generated by the gravitino decay is obtained as
\begin{eqnarray}
\epsilon_B' \equiv
- \sum_i 
\frac{\Delta \Gamma({\psi_\mu \rightarrow  \tilde{q}_{Ri} \bar{q}_{Ri}})}
{\Gamma_{3/2}} {\rm Br}^{\tilde{q}_i}.
\end{eqnarray}

Now, let us consider only nonzero $\lambda_{332}$ as in the
Sec.~\ref{sec:Basym}.  In the large gravitino mass limit, we
get\footnote{In Ref.~\cite{Cline:1990bw} a factor
  $\frac{3}{4}\frac{1}{32\pi^2}$ is given instead of
  $\frac{1}{32\pi^2}$.}
\begin{eqnarray}
\sum_i \Delta \Gamma({\psi_{\mu} \rightarrow \tilde{q}_{Ri} \bar{q}_{Ri}})
 &\simeq&
\frac{1}{32\pi^2}
\frac{|\lambda_{332}|^2{\rm Im}(A_{332}m^*_{3/2})|m_{3/2}|}
{M_P^2},
 \end{eqnarray}
which gives rise to
\begin{eqnarray}
  \sum_i 
\frac{\Delta \Gamma({\psi_{\mu} \rightarrow \tilde{q}_{Ri} \bar{q}_{Ri}})}
  {\Gamma_{3/2}} \simeq
  \frac{3}{260\pi}\frac{|\lambda_{332}|^2{\rm Im}(A_{332}m_{3/2}^*)}{|m_{3/2}|^2}.
\end{eqnarray}
Taking ${\rm Br}^{\tilde{t},\, \tilde{b},\, \tilde{s}}={\rm
  Br}^{\tilde{q}}$, net baryon number generated due to this process is
given by
\begin{eqnarray}
\epsilon_B' &\simeq&
5.6\times 10^{-4}
\left(\frac{|\lambda_{332}|^2\,{\rm Im}
(A_{332} m^*_{3/2})/|m_{3/2}|^2}{-0.1}\right)
 \left(\frac{{\rm Br}^{\tilde{q}}}{0.5}\right).
\label{eq:kappa_{3/2}}
\end{eqnarray}
Then baryon asymmetry directly generated by the gravitino decay to
quark and squark is
\begin{eqnarray}
\frac{n_B}{s}\Big|_{\rm squark} &=& \frac{3T_{3/2}}{4m_{3/2}} \epsilon_B'
\nonumber \\
&\simeq&
8.3\times 10^{-10} \left(\frac{g_\ast}{106.75}\right)^{-1/4}
\left(\frac{m_{3/2}}{10^8~{\rm GeV}}\right)^{1/2}
\left(\frac{\epsilon_B'}{6\times 10^{-4}}\right).
\label{eq:etaB_{3/2}_squark}
\end{eqnarray}
Therefore, it is seen that the baryon number which is resulted in the
processes $\psi_\mu \rightarrow \tilde{q}_{R}\bar{q}_{R}$,
$\tilde{q}_R^* q_R$ has the same order as gluino-mediated one.

Finally let us compare those results with Ref.~\cite{Cline:1990bw}. In
their work, they took the maximum value for CP phase which is allowed
by constraints from neutron EDM experiment at that time. ($d_n \le
10^{-26}~e~{\rm cm}$ was used there.) We have estimated the CP phase
by following the way they described in their paper.  Then taking ${\rm
  Br}^{\tilde{q}_i}=1$ and $|\lambda_{332}|=4 \pi \times 0.1$ as they
did, we get
\begin{eqnarray}
\frac{n_B}{s}&=&
\frac{n_B}{s}\Big|_{\rm gluino}+\frac{n_B}{s}\Big|_{\rm squark}
\nonumber \\ &\simeq&
1.6 \times 10^{-10}
\left(\frac{g_\ast}{10.75}\right)^{-1/4}
\left(\frac{m_{3/2}}{2~{\rm TeV}}\right)^{1/2}.
\end{eqnarray}
This is almost consistent with Ref.~\cite{Cline:1990bw} up to ${\cal
  O}(1)$ factor.

\subsection{Saxion-induced baryogenesis}

The Peccei-Quinn (PQ) mechanism is a plausible solution to the strong
CP problem, and it predicts a pseudo Nambu-Goldstone boson, the
axion~\cite{Peccei:1977hh,QCD-axion}, which acquires a small mass
predominantly from the QCD anomaly.  The axion is stable on
cosmological time scale, thus it is a good candidate for dark matter.
In the supersymmetric extension of the PQ mechanism, its scalar
partner, called saxion, remains relatively light as it acquires a mass
only from the SUSY breaking effects.  In fact, it is known that the
saxion tends to dominate the energy density of the early Universe, and
it plays a similar role as the modulus.  The advantages of using the
saxion as a source of the baryogenesis are two folds. First, it
necessarily couples to the QCD gauge sector in order to solve the
strong CP problem, and therefore naturally decays into gluons and
gluinos. Secondly, the baryogenesis can be more efficient because the
decay temperature will be relatively high for the axion decay constant
$F_a$ is smaller than the GUT scale.

Let us assume that the coherent oscillation of the saxion dominates
the energy density of the Universe. The amount of baryon asymmetry
generated by the saxion decay depends on the reheating temperature
$T_s$ at the saxion decay as well as the branching fraction for the
saxion decay into a gluino pair.  In particular, the latter depends on
the saxion coupling to axion,
\begin{eqnarray}
\frac{\xi}{F_a} s (\partial a)^2,
\end{eqnarray}
where $s$ and $a$ are the saxion and axion, respectively, and $\xi$ is
a model-dependent numerical coefficient.  In order to have a sizable
branching fraction into gluino pair, and also to avoid the
overproduction of axionic dark radiation, $\xi$ must be highly
suppressed below unity since the saxion coupling to the QCD gauge
sector is one-loop suppressed for a given axion decay constant
$F_a$.\footnote{ The branching fraction into the axions can also be
  suppressed by introducing a coupling of the saxion to the Higgs
  fields~\cite{Jeong:2012np} or the right-handed
  neutrinos~\cite{Jeong:2013axf}. The axionic dark radiation has been
  extensively studied in {\it e.g.},
  Refs.~\cite{Chun:2000jr,Ichikawa:2007jv,Jeong:2012np,Jeong:2013axf,Cicoli:2012aq,Higaki:2012ar,Higaki:2013lra}.
} As the precise value of $\xi$ depends on how the PQ scalars are
stabilized, let us see the axion models in some detail.

The axion models can be categorized according to whether the PQ
symmetry is realized linearly or non-linearly.  For the former case,
which is sometimes dubbed a field-theoretic axion model, the axion
decay constant $F_a$ ranges from an intermediate scale up to the GUT
scale. The saxion coupling with axion depends on the stabilization
mechanism. In a model with a single PQ scalar, $\xi$ is of order
unity.  One way to suppress $\xi$ is to consider a model with two PQ
scalars $S_1$ and $S_2$, which have an opposite PQ charge each other,
and a U(1)$_{\rm PQ}$ singlet $\Sigma$,
\begin{eqnarray}
\label{axion-model}
W_{\rm PQ} = \lambda \Sigma(S_1 S_2 - \mu^2),
\end{eqnarray}
so that the PQ symmetry is broken along the $F$-flat direction
$S_1S_2=\mu^2$, which is lifted by SUSY breaking effects.  In this
model, the axion decay constant is given by $\mu$.  If $S_1$ and $S_2$
obtain the soft scalar masses of a similar size, then $\xi$ is
suppressed as $\xi^2\sim \Delta m^2_S/m^2_S$, where $m^2_S$ is the
typical size of their soft masses and $\Delta m^2_S$ is the mass
splitting between them~\cite{Jeong:2013axf}.

The latter case, which is called a string axion model, on the other
hand, corresponds to the case where the axionic shift symmetry of some
moduli remains unbroken (except for the QCD anomaly).  The axion decay
constant is then typically around $M_P/8\pi^2$ unless the K\"ahler
metric of the saxion is hierarchically smaller than unity at the
vacuum.  Thus, in this case, one needs to assume a small initial
misalignment of the axion in order for the axion relic energy density
not to overclose the Universe.  On the other hand, the saxion coupling
to axion is typically suppressed as $\xi \sim 1/8\pi^2$, and thus the
branching fraction of the saxion decay into gluino pair is naturally
sizable and is comparable to that into the axions~\cite{Higaki:2013lra}.
This can be understood by noting that the coupling to the QCD gauge
sector arises at tree level in this case.

We note that the KKLT mechanism provides a natural framework to
implement the saxion-induced baryogenesis both for the field-theoretic
and the string axion models.  First, the CP phase is obtained in the
same way as explained in Sec.~\ref{sec:2}. Note that the $F$-term in
the axion multiplet does not induce a new CP phase in the MSSM soft
terms because the soft terms generating a saxion potential depend on
the moduli $F$-terms and the gravitino mass.  Besides, the suppression
for $\xi$ can be also implemented in both models with the KKLT
scenario.  For the axion model described in Eq.~(\ref{axion-model}),
let us assume that the Yukawa couplings of $S_1$ and $S_2$ are small,
and the PQ scalars $S_1$ and $S_2$ have the same modulus dependence,
{\it i.e.}, the same modular weight.  Since the symmetry under
interchanging between $S_1$ and $S_2$ is a good symmetry, then one can
naturally achieve $\Delta m^2_S \ll m^2_S$, suppressing the saxion
coupling to the axion.  On the other hand, in the string axion model,
the QCD axion can arise from a K\"ahler modulus in the KKLT scenario
with multiple K\"ahler moduli, if the axionic shift symmetry for the
K\"ahler modulus is not broken
explicitly.~\cite{Choi:2006za,Conlon:2006tq}.  This modulus is
stabilized with a mass of about $\sqrt2 m_{3/2}$ through K\"ahler
mixing with other moduli, which are stabilized by the non-perturbative
superpotential as in the original KKLT.

In either field-theoretic axion model or string axion model, the
modulus $X$ may dominate the Universe before the saxion-dominated
Universe. This is because the saxion is much lighter than the
modulus. Although the modulus decay generates the baryon asymmetry as
we have described, the asymmetry is washed out by the saxion decay
later. However, the saxion decay can generate baryon asymmetry.

Now let us estimate the baryon asymmetry produced by the saxion.  For
simplicity we consider the case where the axion production from the
saxion decays is negligible, and the saxion dominantly decays to a
gluon pair and a gluino pair. Including the decay into the axions does
not affect our results as long as $|\xi| \lesssim 1/8\pi^2$.

The saxion couplings to gluon and gluino are given (in two-component
notation) by
\begin{eqnarray}
{\cal L}_s=
\frac{\alpha_s}{8\pi} \frac{s}{F_a}G_{\mu\nu}G^{\mu\nu}
- \frac{\alpha_s}{8\pi} \frac{m_ss}{F_a}
 (\kappa \tilde{g}\tilde{g}+{\rm h.c.}),
\end{eqnarray}
where $G_{\mu\nu}$ is the gluon field strength and $m_{s}$ is the
saxion mass, and the order unity constant $\kappa$ is determined by
how the saxion is stabilized.  The above couplings mediate the saxion
decay into a gluon pair and a gluino pair with
\begin{eqnarray}
&&
\Gamma(s \rightarrow gg)=\frac{\alpha_s}{32\pi^3}\frac{m_{s}^3}{F_a^2},
\\ &&
\Gamma(s \rightarrow \tilde{g}\tilde{g})=
\frac{\alpha_s}{32\pi^3}\frac{m_{s}^3}{F_a^2}
\left[1-\frac{4M_{\tilde{g}}^2}{m_s^2}\right]
\left[|\kappa|^2\left(1-\frac{M_{\tilde{g}}^2}{2m_s^2}\right)
-2{\rm Re}(\kappa^2)\frac{M_{\tilde{g}}^2}{m_s^2}\right].
\end{eqnarray}
Then the total decay rate of the saxion is given as
\begin{eqnarray}
\Gamma_s \simeq
\frac{\alpha_s}{32\pi^3}\frac{m_{s}^3}{F_a^2}
\left(1+|\kappa|^2\right),
\end{eqnarray}
neglecting the gluino mass in the final state.  Consequently the
branching ratio for the process $s \rightarrow \tilde{g}\tilde{g}$ is
obtained as
\begin{eqnarray}
{\rm Br}(s \rightarrow \tilde{g}\tilde{g})\simeq
\frac{|\kappa|^2}{1+|\kappa|^2}.
\label{eq:Br_saxion_gluino}
\end{eqnarray}
On the contrary, ${\rm Br}(s \rightarrow \tilde{g}\tilde{g})\simeq
(1+m_s^2/8M_{\tilde{g}}^2)^{-1}$ was given in
Ref.~\cite{Mollerach:1991mu}. This result indicates a chiral
suppression due to the gluino mass since ${\rm Br}(s \rightarrow
\tilde{g}\tilde{g}) \rightarrow 0$ when $M_{\tilde{g}}\rightarrow
0$. As it is seen, however, there is no such chiral suppression in
Eq.~(\ref{eq:Br_saxion_gluino}), which is already pointed out by
Ref.~\cite{Endo:2006ix}.  Since $|\kappa|$ is order unity, the saxion
can dominantly decay into a gluino pair.  This makes the baryogenesis by
the saxion decay more efficient than originally considered in
Ref.~\cite{Mollerach:1991mu}.

After the gluino production, the same story follows as in the modulus
case.  In the string axion models, the reheating temperature is
similar to the modulus case, while it can be higher in field-theoretic
axion models if the axion decay constant is smaller than the GUT
scale.  The reheating temperature is given by
\begin{eqnarray}
T_s \simeq
0.73~{\rm GeV}
\left(\frac{g_\ast}{10.75}\right)^{-1/4}
\left(\frac{10^{16}~{\rm GeV}}{F_a}\right)
\left(\frac{m_s}{10^5~{\rm GeV}}\right)^{3/2},
\label{eq:TR_saxion}
\end{eqnarray}
where we have taken $\kappa=1$.  The resultant asymmetric yield is
estimated by
\begin{eqnarray}
\frac{n_B}{s} &=&
 \frac{3T_s}{4m_s} 2{\rm Br}(s\rightarrow \tilde{g}\tilde{g})\epsilon_B
\nonumber \\ &\simeq&
5.4\times 10^{-10}
\left(\frac{g_\ast}{10.75}\right)^{-1/4}
\left(\frac{m_s}{10^5~{\rm GeV}}\right)^{1/2}
\left(\frac{10^{16}~{\rm GeV}}{F_a}\right)
\left(\frac{{\rm Br}(s\rightarrow \tilde g\tilde g)}
{1/2}\right)
\left(\frac{\epsilon_B}{10^{-3}}\right).
\nonumber \\
\end{eqnarray}
Thus the right amount of baryon asymmetry can be generated for the
saxion mass around $10^3-10^5~{\rm GeV}$.  On top of that, as
mentioned above, the axion produced by the misalignment mechanism
contributes to the cold dark matter.  This fact is especially
important because the lightest sparticle is no longer stable due to
the R-parity violation.

%%%%%%%%%%%%%%%%%%%%%%%%%%%%%%%%%
\section{ Conclusions and Discussion}
\label{sec:5}
%%%%%%%%%%%%%%%%%%%%%%%%%%%%%%%%%
In this paper we have studied a baryogenesis induced by late-decaying
moduli, and examined its implications for the moduli stabilization and
the SUSY breaking scale.  If the branching fraction of the modulus
into a gluino pair is sizable, the right amount of baryon asymmetry
can be generated through CP violating decay of gluino into quark and
squark followed by baryon-number violating squark decays.  We have
shown that a natural framework realizing the baryogenesis is provided
by the KKLT-type moduli stabilization since sufficient CP violation is
obtained from mixed modulus-anomaly mediated SUSY breaking in the
presence of two or more non-perturbative terms in the modulus
superpotential.  Successful baryogenesis is possible for the gravitino
mass around $10^5$--$10^7$~GeV or equivalently the soft SUSY breaking
mass $m_{\rm soft}\sim 10^3$--$10^5$~GeV (or heavier for a suppressed
CP phase or effective CP violation parameter).  Such low SUSY breaking
scale can be probed directly at the collider experiments, dinucleon
decay search, and electric dipole moments of neutron and electron.  We
also found that similar baryogenesis works successfully in other
scenarios where the saxion or the gravitino dominates the Universe.

Lastly let us discuss baryogenesis in a couple of other moduli
stabilization and SUSY breaking scenarios.  We have mainly focused on
the mixed modulus-anomaly mediation so far, assuming a sequestered
uplifting. If the uplifting is not sequestered, sfermion masses are
generically heavier than gaugino masses, then the gluino decay into
quark and squark will be kinematically forbidden.  Even in such case,
however, the gravitino produced by the modulus decay will be able to
generate baryon asymmetry.

The large volume scenario~\cite{Conlon:2005ki} can also lead to CP
violating soft terms if one takes the same form of the superpotential
for a small 4-cycle modulus as in \EQ{W-model}.  However, one cannot
simply apply the moduli-induced baryogenesis discussed in the present
paper to the large volume scenario because the overall volume modulus,
which has a small mass compared to other moduli, does not appear in
the visible gauge kinetic function and thus weakly couples to gluino.
In addition, the modulus decay into a pair of gravitinos is
kinematically forbidden and the decay into Higgsinos is suppressed due
to the approximate no-scale structure even in the presence of the
Giudice-Masiero term~\cite{Higaki:2012ar}. Those fact indicate much
smaller amount of gluino produced by modulus decay, which suppress the
resultant baryon number.

Regarding the moduli stabilization, we note that there is a general
tension between moduli stabilization by non-perturbative effects and
chirality~\cite{Blumenhagen:2007sm}.  To be specific, (a combination
of) the moduli $X_L$ appearing in the gauge kinetic function of the
visible gauge sector may not be stabilized by instantonic exponential
terms. The saxion in the string axion model corresponds to this case,
and as discussed before, it can be stabilized by another
way~\cite{Choi:2006za,Conlon:2006tq} and the saxion-induced
baryogenesis works.  Alternatively, we may consider a modulus field
which is stabilized {\it a la} KKLT, while not appearing in the SM
gauge kinetic function. Such modulus field can still decay into gauge
bosons and gauginos through the kinetic mixing with $X_L$. Thus the
baryon asymmetry will be generated by the subsequent gluino decay,
although it may be diluted by the decay of lighter moduli $X_L$ to
some extent.

Similar baryogenesis can be also realized in gravity mediation.  In
the gravity-mediated SUSY breaking, the MSSM gaugino masses are
generated by Planck-scale suppressed interactions with an elementary
gauge singlet field called the Polonyi field.  Since it is a singlet
under any symmetry of the theory, there is no special point in its
field space. Therefore, during inflation, the Polonyi field is
generically deviated from the true minimum, as the effective potential
for the Polonyi field during inflation is deformed by its
gravitational couplings with the inflaton.  After inflation, the
Polonyi starts to oscillate about the minimum with an amplitude of
order the Planck scale, and soon dominates the energy density of the
Universe.  The Polonyi field will decay into the MSSM gauge sector,
and thus the baryogenesis by the Polonyi decay will be possible. In
particular, a sizable A-term as well as large CP phases are naturally
generated in the gravity mediation, which nicely fits with the current
scenario. The resultant baryon asymmetry is similar to (\ref{eq:etaB})
if the modulus mass is replaced with the Polonyi mass, which is order
of the gravitino mass.

%%%%%%%%%%%%%%%%%%%%%%%%%%%%%%%%%%%%%
\section*{Acknowledgment}
%%%%%%%%%%%%%%%%%%%%%%%%%%%%%%%%%%%%%
This work is supported by Grant-in-Aid for Scientific Research on
Innovative Areas (No.24111702 , No. 21111006, and No.23104008) [FT],
Scientific Research (A) (No. 22244030 and No.21244033) [FT], and JSPS
Grant-in-Aid for Young Scientists (B) (No. 24740135) [FT], Inoue
Foundation for Science, and by World Premier International Center
Initiative (WPI Program), MEXT, Japan [FT]. This publication is partly
funded by the Gordon and Betty Moore Foundation through Grant GBMF
\#776 to the Caltech Moore Center for Theoretical Cosmology and
Physics. The work is also supported by the U.S. Department of Energy
under Contract No. DE-FG02-92ER40701 [KI].

\end{document}